\documentclass{pasj00}
\draft
\usepackage{natbib}
\bibpunct{(}{)}{;}{a}{}{,}

\begin{document}
\SetRunningHead{S. Kamio}{Jets in Coronal Hole}
\Received{2007/05/31}%{yyyy/mm/dd}
\Accepted{2007/--/--}%{yyyy/mm/dd}

\title{Velocity Structure of Jets in Coronal Hole}

%%% begin:list of authors
% Do NOT capitalize all letters in "textsc".
\author{Suguru \textsc{Kamio}, Hirohisa \textsc{Hara}, Tetsuya \textsc{Watanabe}}
\affil{Hinode Science Center, Solar Physics Division, National Astronomical Observatory of Japan, 2-21-1 Osawa, Mitaka, 181-8588 Japan}
\email{suguru.kamio@nao.ac.jp}

\author{Keiichi \textsc{Matsuzaki}}
\affil{Institute of Space and Astronautical Science, Japan Aerospace Exploration Agency, 3-1-1 Yoshinodai, Sagamihara, Kanagawa, 229-8510 Japan}
\author{Kazunari \textsc{Shibata}}
\affil{Kwasan and Hida Observatories, Graduate School of Science, Kyoto University, Yamashina, Kyoto, 607-8471 Japan}
\author{Len \textsc{Culhane}}
\affil{Mullard Space Science Laboratory, Department of Space and Climate Physics,University College London, Holmbury St. Mary, Dorking, Surrey RH5 6NT UK}
\and
\author{Harry \textsc{Warren}}
\affil{Solar Terrestrial Relationships Branch, Naval Research Laboratory, 4555 Overlook Avenue, SW Washington, DC 20375 USA}

%%% end:list of authors

%%% Please use the following style in case that sorting by 
%%% affilation is impossible. 
%
% \author{%
%   D-Firstname \textsc{D-Familyname}\altaffilmark{1}
%   E-Firstname \textsc{E-Familyname}\altaffilmark{1,2}
%   and
%   F-Firstname \textsc{F-Familyname}\altaffilmark{2}}
% \altaffiltext{1}{Address of Institute}
% \email{ddddd@xxx.xxx.xx.xx}
% \email{eeeee@xxx.xxx.xx.xx}
% \altaffiltext{2}{Address of Institute}

%% `\KeyWords{}' always has to be placed before `\maketitle'.
 %Do NOT move this preamble from here!

\maketitle

\begin{abstract}
Velocity structures of jets in a coronal hole have been derived
for the first time.
{\it Hinode} observations revealed the existence of many bright
points in coronal holes. They are loop-shaped and sometimes
associated with coronal jets.
Spectra obtained with the Extreme ultraviolet Imaging Spectrometer
(EIS) on board {\it Hinode} are analyzed to infer Doppler velocity
of bright loops and jets in a coronal hole of the north polar region.
Elongated jets above bright loops are found to be
blue-shifted by 30 km s$^{-1}$ at maximum, while foot points of
bright loops are red-shifted.
Blue-shifts detected in coronal jets are interpreted as
upflows produced by magnetic reconnection between emerging flux and
the ambient field in the coronal hole.
\end{abstract}

\section{Introduction}
Coronal holes are dark regions of the corona
seen in soft X-ray or coronal spectral lines.
In a coronal hole, there is open magnetic field
where solar wind is flowing out, while bright corona
of the surrounding area is composed of closed field.
It is obvious that magnetic field plays an important
role in forming varying structure of corona, but
the mechanism responsible for energy and mass
supply to the corona remains a mystery.
Study of the dynamics in coronal holes is, therefore,
crucial for understanding coronal heating and 
the acceleration of the solar wind.

\cite{shimojo1996} and \cite{shimojo2000} derived
physical parameters of X-ray jets using
the soft X-ray telescope \citep{tsuneta1991}
on board {\it Yohkoh}
spacecraft \citep{ogawara1991}.
They claimed that the observed properties of
X-ray jets were consistent with the theory, in which
jets were produced by evaporation flows due to
reconnection heating.

\cite{moses1997} studied coronal jets in a sequence of
195\AA~ images obtained with EIT on {\it SOHO}.
They found apparent rising velocity of the jets is 100 -- 400 km s$^{-1}$.

But the measurement of Doppler velocity of coronal jets
is needed to firmly establish the model, because
their analysis is based on the apparent motion of the
jets which may differ from the actual plasma motion.

\cite{wang1998} identified white-light jets in
2 -- 6 R$_{\odot}$ range of LASCO images,
which corresponds to EUV jets observed with EIT.
They were rooted in bright points
The leading edge of white-light jets propagated with speeds of
400 -- 1000 km s$^{-1}$.

Coronal plume is a stationary structure extending
from coronal holes observed in the EUV and coronagraphs in
visible light.
\cite{deforest1997} traced plume structures from the surface to
15 R$_{\odot}$ by combining EIT and coronagraph data.
While coronal jets are transient phenomena,
plume structures are stable, at least for 24 hours.
The study of temporal variation is important to
distinguish jets and plumes.

The Extreme ultraviolet Imaging Spectrometer (EIS; \citealt{culhane2007})
on board {\it Hinode} \citep{kosugi2007} is capable of obtaining
EUV spectra with high resolution and high efficiency.
It is designed to derive properties of the solar corona,
such as temperature, line-of-sight velocity, and
density, through spectroscopic observation.
Although features in coronal hole are dark in coronal
spectral lines, the high throughput of the EIS allows us
to determine Doppler shifts in coronal holes.
In the present work, we studied the velocity structure
of bright points in a coronal hole in the north polar region.
The EIS observations are described in \S 2.
The data reduction procedure and the method of data analysis are
shown in \S 3.
Results are presented in \S 4 and their interpretation
is discussed in \S 5.
\section{Observation}

\begin{figure*}
  \begin{center}
    \FigureFile(160mm,80mm){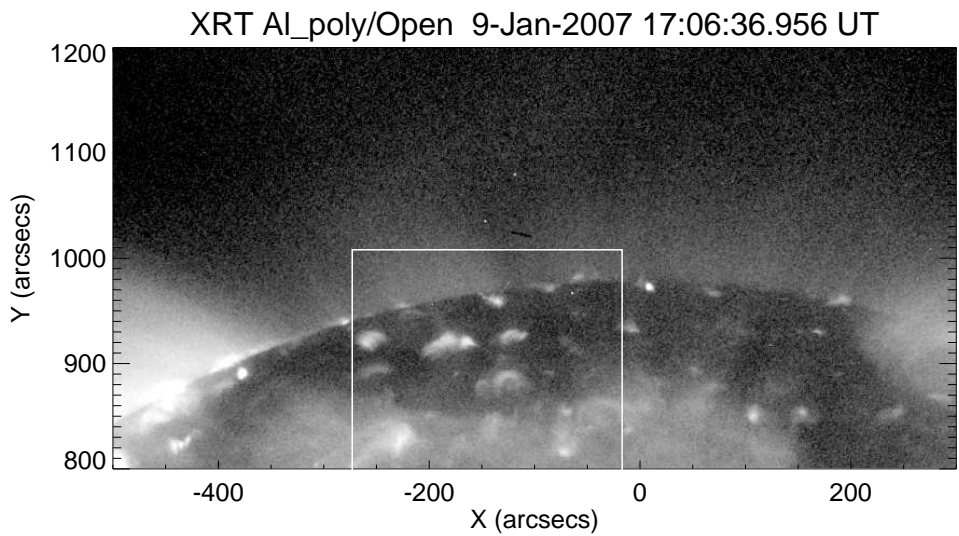}
  \end{center}
  \caption{X-ray jets in the polar coronal hole observed with XRT.
The overlaying box indicates the EIS field of view.
The bottom of the EIS field of view is outside that of XRT.}\label{fig:xrt}
\end{figure*}

A coronal hole in the north polar region was observed by
{\it Hinode} on 9 January 2007. Figure \ref{fig:xrt} shows
an X-ray image obtained with the X-Ray Telescope
(XRT; \citealt{golub2007}).
Many bright points are found even in a coronal hole.
They are loop-shaped and X-ray jets are occasionally
formed from bright points.

One raster scan was performed with the EIS from 16:10 -- 18:24 UT.
The EIS field of view is overlaid on Fig. \ref{fig:xrt}.
EIS has two spectral ranges, 170--210\AA~ and 250--290\AA~, and
the spectral windows of Fe X 184\AA, Fe VIII 185,
Fe XI 188\AA, Ca XVII 192\AA, Fe XII 195\AA,
Fe XIII 202\AA, He II 256\AA, Fe XIV 274\AA,
and Fe XV 284\AA~ were selected and recorded. 
The slit of 1'' width was selected to determine
the Doppler shift of spectral lines.
The raster swept a 256''$\times$256'' area with
a 30 sec exposure at each slit position.
Since this observation was aimed at studying spatial distribution
of coronal hole in a large area, the temporal evolution of 
coronal features is not analyzed in the present paper.

\begin{figure*}
  \begin{center}
    \FigureFile(160mm,160mm){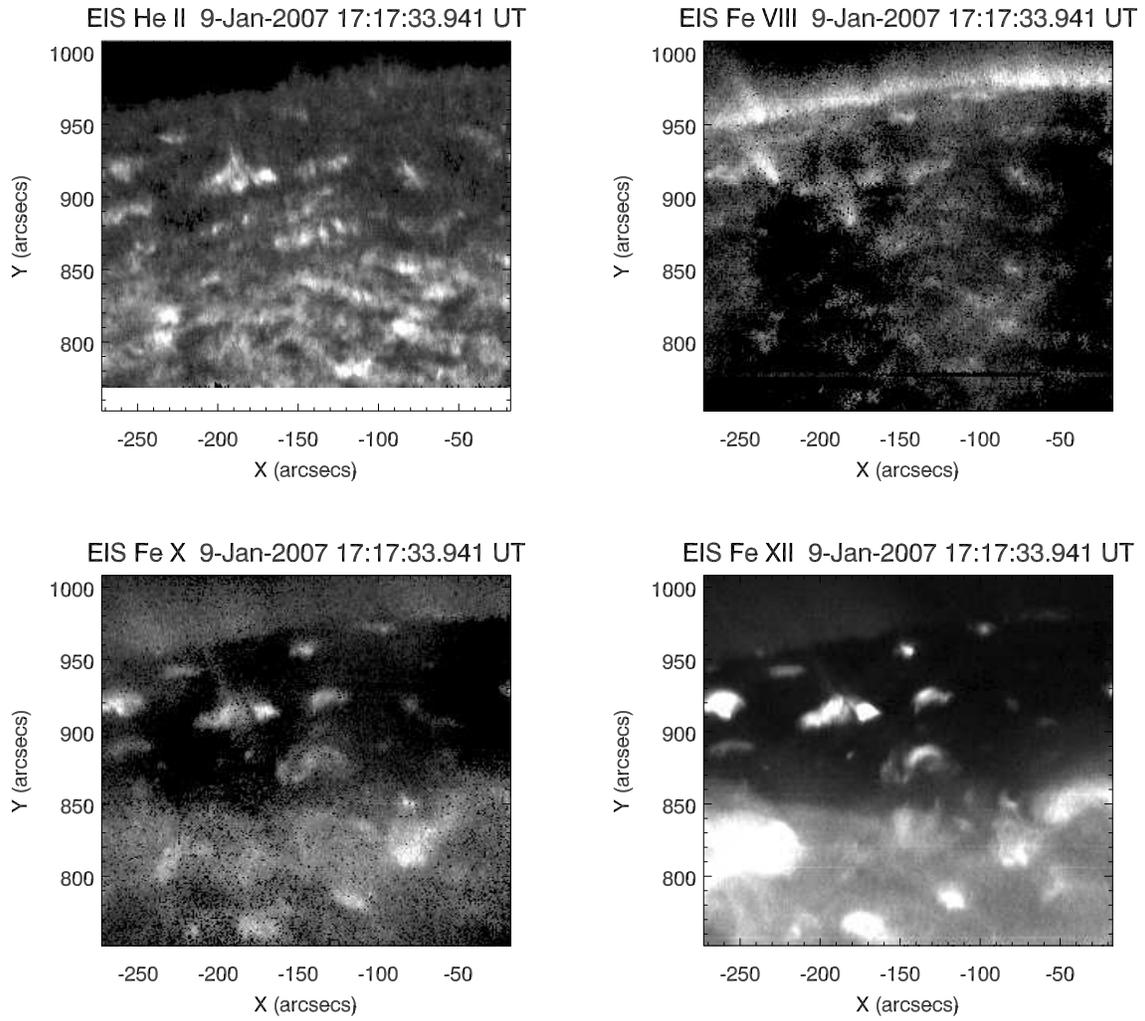}
  \end{center}
  \caption{Intensity maps in selected lines from the EIS data set.
The coronal hole boundary is around $Y=850$, and upper regions correspond to part of the coronal hole in Fe {\textsc XII} 195\AA~
($\log {\rm T}_{\rm e} = 6.1$) and Fe {\textsc X} 184\AA
($\log {\rm T}_{\rm e} = 6.0$).

The coronal hole is not noticeable in Fe {\textsc VIII} 185\AA~ or
He {\textsc II} 256\AA~. 
The north limb and the corona above the limb are
found near the upper end of the field.}\label{fig:snap}
\end{figure*}

Figure \ref{fig:snap} presents intensity maps in
selected lines.
The field of view covered a part of the coronal hole
and surrounding quiet region in the north polar region.
In the Fig. \ref{fig:snap}, the coronal hole boundary is
located at about $Y = 850''$.
In this data set, the corona above the limb was also observed
near the upper end of the field.
In prominent Fe {\textsc XII} 195\AA~ line ($\log {\rm T}_{\rm e} = 6.1$),
data numbers sufficient for spectrum fitting were obtained
in the dark coronal hole, which allowed us to study the Doppler
velocity structure over the entire field.
XRT obtained a sequence of images at the same time, showing the apparent
motion of jets elongating from bright points in the coronal hole.

\section{Data Reduction}
Basic reduction of the EIS data was carried out by
standard procedures provided in Solar Software (SSW).
The EIS data consist of series of EUV spectra at each slit
position in the raster scan.
Dark current subtraction and flat field correction
were applied to the raw data.
Spikes caused by cosmic-rays and hot pixels were
marked as invalid data and ignored in the following
process.

Single Gaussian component fitting was performed
with calibrated spectra to determine the center
position, amplitude, and width of the spectral lines.
If the data count was not enough to perform
spectral fitting, the pixel was treated as no
signal. Because no distinct case of a two component
spectral line was found in this data, single component
fitting gives reliable results.

In order to derive Doppler shift of the spectrum,
a wavelength reference must be defined as the EIS
does not have an absolute wavelength reference.
Line center positions determined from spectral fitting
indicate cyclic modulation, since the EIS instrument
experiences temperature variation in sync with the
spacecraft orbit.
It leads to an orbital variation of the line center
position ranging to 70 km s$^{-1}$
in Doppler velocity in peak-to-peak.

Since this data set covers long range in the slit direction
(256'' height), local coronal structures are smoothed
out by averaging the line center positions along the slit.
The orbital variation of the spectrum position
can be computed over the raster and subtracted from the
line shift to determine true Doppler shift.
The result is the velocity with respect to the average
over the field.
Angle of slit tilt, which results in artificial
wavelength shift of the spectrum, is determined by
using quiet region observation at disk center
at 16:02 UT and 21:43UT, before and after this coronal
hole observation.
The slit tilt is removed in the
Doppler velocity measurement.

The mean velocity is not zero in coronal temperature lines.
\cite{sandlin1977} determined a net blueshift of 6 km s$^{-1}$
in Fe {\textsc XII} from the Sun.
\cite{peter1999} found a temperature dependence of
Doppler shifts which is consistent with the result from
\cite{sandlin1977}.
Assuming that non-radial velocities are canceled in the corona,
the Doppler shift outside the limb must be zero.
In the present work, the off-limb area excluding jets is regarded
as the velocity reference.
This method may introduce ambiguity in the velocity,
but the error must be smaller than the net blueshift of
6 km s$^{-1}$ found by \cite{sandlin1977}.

The Doppler velocity is only derived for Fe {\textsc XII}
line, which has a large data count to perform
spectrum fitting for the most of the field.
In other spectral lines, spectrum fitting can be done for
a limited portion of the data and it is hard to define a
reference position in spectra.

\section{Results}

Figure \ref{fig:snap} shows intensity maps in He {\textsc II},
Fe {\textsc VIII}, Fe {\textsc X}, and Fe {\textsc XII}.
In the Fe {\textsc XII} panel showing coronal plasma
($\log {\rm T}_{\rm e} = 6.1$),
several bright points are found within the dark coronal hole.
Some of them show loop-shaped structure with faint threads
stretching above the loop.
Similar features are found in Fe {\textsc X}, it is
degraded because of the low data count.
At dark pixels, spectrum fitting was not performed
because of low signal to noise ratio.
Fe {\textsc VIII} represents lower temperature plasma
($\log {\rm T}_{\rm e} = 5.6$),
and the boundary of the coronal hole is not obvious.
Comparison with Fe {\textsc XII} indicates
that the foot points of bright loops in Fe {\textsc XII}
are bright in Fe {\textsc VIII}, implying
lower temperature in the foot points.
As the Fe {\textsc VIII} mainly represent transition region
features, it outlines the lower end of the corona and
is seen as a bright layer above the limb.
He {\textsc II} shows the lowest temperature available for EIS
($\log {\rm T}_{\rm e} = 4.7$).
There are bright features associated with the coronal bright
points, but they may be contaminated by a Si {\textsc X} line
blended in the red wing of He {\textsc II} spectrum.

\begin{figure}
  \begin{center}
    \FigureFile(80mm,80mm){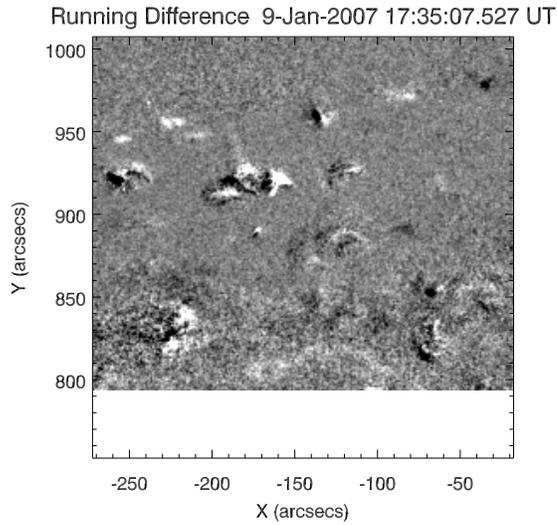}
  \end{center}
  \caption{Running difference of XRT data obtained at
17:23 UT and 17:35 UT. White indicates intensity
increase in the X-ray. The FOV is the same as in Fig. \ref{fig:snap}.
}\label{fig:diff}
\end{figure}

Figure \ref{fig:diff} shows running difference
of XRT data indicating an outward propagating jet (X = -180).
A sequence of XRT data allow us to find the
temporal change of the jet and to distinguish
it from stationary plumes.
The EIS slit scanned the jet at about 17:40 UT
when the jet was growing in XRT.

\begin{figure}
  \begin{center}
    \FigureFile(80mm,80mm){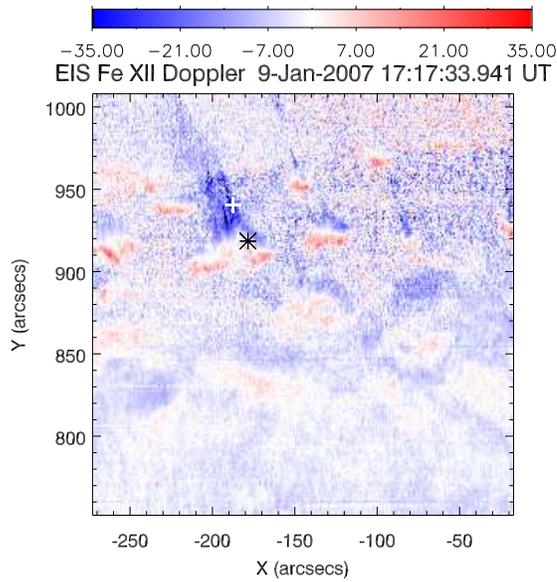}
  \end{center}
  \caption{Doppler velocity inferred from Fe {\textsc XII} spectra.
Color scale represents the velocity from -35 km s$^{-1}$ to +35 km s$^{-1}$.
Positive and negative velocities correspond to red and blue shifts,
respectively. Cross and asterisk symbols indicate the location
of spectra in Fig. \ref{fig:spectrum}.
}\label{fig:vel}
\end{figure}

Doppler velocity derived from Fe {\textsc XII} line
is shown in Fig. \ref{fig:vel}.
Comparing with the intensity map in Fig. \ref{fig:snap},
bright points in coronal hole have noticeable Doppler shifts.
In this data, loop foot points of bright loops are red-shifted
by 15 km s$^{-1}$.
This is larger than the velocity ambiguity of 6 km s$^{-1}$,
hence it is real red-shift suggesting downflow in the loop.

The magnitude of red-shift decreases from the bottom of the loop
to the top.
The velocity becomes zero at the asterisk symbol in Fig.
\ref{fig:vel}.
A blue-shifted feature is stretching above the bright point.
The maximum velocity of -30 km s$^{-1}$ is attained at
the cross symbol in Fig. \ref{fig:vel}, where a faint thread
extends upward in Fig. \ref{fig:snap}.

There are other bright points showing similar structure,
but their Doppler amplitude are smaller.
It is interesting to note that the direction of these blue-shifted
feature above the bright points are close to the radial direction,
taking heliocentric coordinates into account.
There are bright points outside the coronal hole, but their
velocities are much smaller and no stretching feature
is found.

\begin{figure}
  \begin{center}
    \FigureFile(80mm,80mm){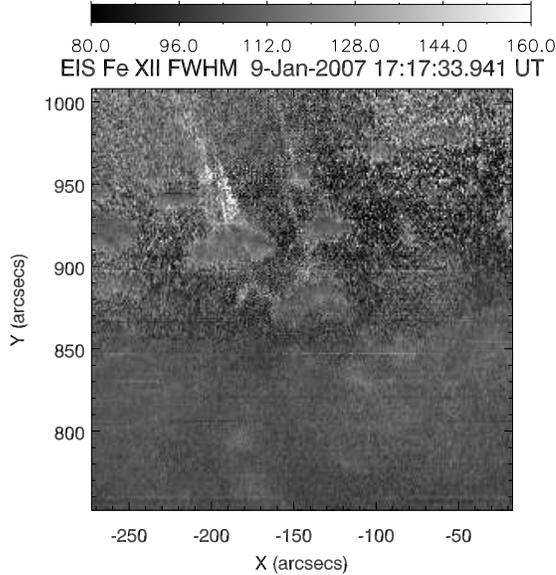}
  \end{center}
  \caption{Distribution of full width at half maximum of Fe {\textsc XII}
spectra.
The instrumental width is not compensated in this figure.
Gray scale spans from 80 km s$^{-1}$ to 160 km s$^{-1}$ in
Doppler velocity scale.}\label{fig:fwhm}
\end{figure}

Figure \ref{fig:fwhm} presents the full width at half maximum in Fe
{\textsc XII}.
Since the instrument width is not compensated, it shows the distribution of
relative line width.
A significant line broadening is detected in the elongated
feature above the bright point.
The maximum line width enhancement is about 30 km s$^{-1}$ there.
No significant change in line width is found outside
the coronal hole.

\begin{figure}
  \begin{center}
    \FigureFile(80mm,80mm){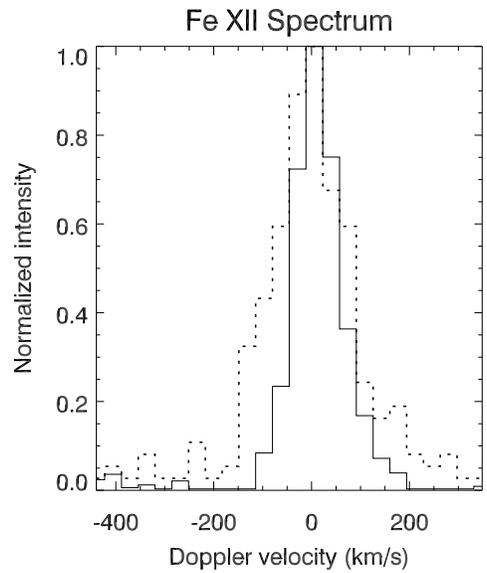}
  \end{center}
  \caption{Solid and dotted lines present Fe {\textsc XII} spectra at asterisk and cross symbols in Fig. \ref{fig:vel}.}
\label{fig:spectrum}
\end{figure}

Raw spectra of Fe {\textsc XII} are presented in Fig. \ref{fig:spectrum}.
The solid line shows the spectrum at the asterisk symbol
in Fig. \ref{fig:vel}, where no significant
line width enhancement is found.
The dotted line spectrum obtained at cross symbol in Fig. \ref{fig:vel}
indicates enhancement in the blue wing.
It is clear that the line broadening in Fig. \ref{fig:fwhm}
is due to the enhancement in the blue wing.

\begin{figure}
  \begin{center}
    \FigureFile(80mm,80mm){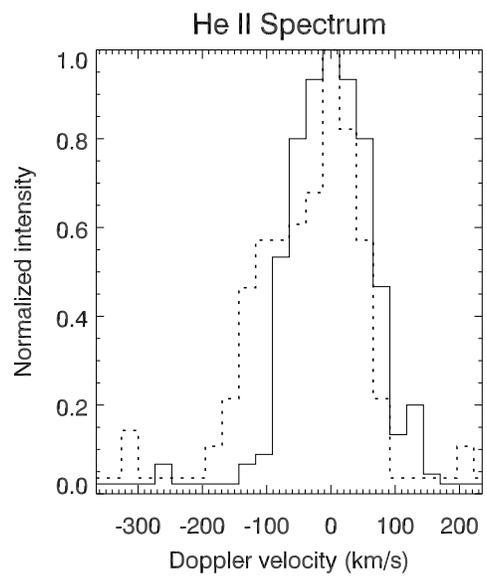}
  \end{center}
  \caption{The same as Fig. \ref{fig:spectrum} but for He {\textsc II} spectra at the same positions.}\label{fig:spectrum_he}
\end{figure}

Figure \ref{fig:spectrum_he} shows He {\textsc II} spectra.
Although the red wing of the spectra may be contaminated
by Si {\textsc X}, enhancement in the blue wing
is attributed to He {\textsc II}.
Broadenings in Fe {\textsc XII} and He {\textsc II}
suggest an explosive nature of the jet.

\section{Discussion}
The intensity maps from EIS clearly show that bright points in the
coronal hole have loop structures at coronal temperature.
Some of them are accompanied by faint elongated features
stretching higher up into the corona.
Doppler velocity measurement revealed blue-shifted
features above bright points.
Assuming that these features are expanding radially,
they imply an uplifting plasma above the bright point.
Comparison with XRT images (Figs. \ref{fig:xrt} and \ref{fig:diff})
indicates that these features are co-spatial with X-ray jets.
\cite{shimojo1996} and \cite{shimojo2000} determined physical parameters
of X-ray jets from {\it Yohkoh} SXT observations.
Their statistical study shows that the apparent velocities of X-ray jets
are 10 -- $10^3$ km s$^{-1}$ with an average of 200 km s$^{-1}$.
Taking the projection effect into account, the measured velocity of
30 km s$^{-1}$ in line-of-sight direction suggests that the actual
velocity of the jets could be one-order of magnitude larger,
which would fit into the range of apparent velocities.

The present observation shows that the bright points in the coronal
hole are red-shifted in their the foot points.
The amount of red-shift decreases and crosses zero-velocity
near the top of bright loop.
Blue-shifted feature above the bright point can be
interpreted as jet, since it is transient
(Fig. \ref{fig:diff}). Spectra broadenings also indicate
explosive nature of the jet.
(Figs. \ref{fig:spectrum} and \ref{fig:spectrum_he}).
\cite{wilhelm2002} analyzed ultraviolet spectrum of jets and
found that the outer wings reached +230 km s$^{-1}$ -- -230 km s$^{-1}$
in Ne {\textsc VIII} ($\log {\rm T}_{\rm e} = 5.8$)
 and +210 km s$^{-1}$ -- -280 km s$^{-1}$
in C {\textsc IV} ($\log {\rm T}_{\rm e} = 5.0$).
In the present paper, enhancements in the blue wings of
Fe {\textsc XII} and He {\textsc II} are found.
They are interpreted as upward motion of the jet in
the corona  ($\log {\rm T}_{\rm e} = 6.1$) and transition region 
($\log {\rm T}_{\rm e} = 4.7$).

\cite{shibata1992} suggested a model of X-ray jet formation,
in which emerging flux collides and reconnects with
pre-existing field.
According to his model, a bright loop is formed by
reconnected field near the foot point of the jets.
The observed loop structures (Fig. \ref{fig:xrt} and \ref{fig:snap})
are thought to be the reconnected fields.
In addition, upward motion is expected along the ambient field line.
\cite{shimojo2001} simulated the evolution of X-ray jets and suggested
that they are evaporation flow produced by magnetic reconnection
taking place near the foot points of the loop.
The strong blue-shift determined in Fig. \ref{fig:vel} is interpreted
as line-of-sight component of upward motion of the jet.
But the foot points are red-shifted by about 15 km s$^{-1}$,
which exceeds the velocity ambiguity of 6 km s$^{-1}$.
If they are not red-shifted, then average velocity of quiet region
outside coronal hole (region of Y $<$ 850'' in Fig. \ref{fig:vel})
must be blue-shifted by 20 km s$^{-1}$ or greater, which is not
realistic.
The observed red-shifts are not consistent with chromospheric
evaporation, since evaporation flow is expected to be an upflow
at the coronal temperature of Fe {\textsc XII}. 
The red-shifts can be reconnected loops moving {\bf downward}
from the reconnection point, but present observation indicates
that all foot points of bright points in coronal hole are
red-shifted.
A more detailed model is needed to explain the observation.

On the other hand, the bright features in the quiet region
outside the coronal hole do not show significant
velocity variation.
The difference in velocity structure reflects the
different field connectivities inside and outside
the coronal hole.
There are ambient open fields in the coronal hole, which
can reconnect with emerging flux to form jets
stretching high to the corona
, but outside fields are closed and
elongated jets are rarely formed.

\section{Summary and Conclusions}
We studied the properties of jets in a coronal hole by using
a raster scan of EIS on board {\it Hinode}.
The high resolution and sensitivity of EIS allowed
us to study the Doppler velocity of coronal jets.
Our observation showed that bright points
in coronal holes are loop-shaped at coronal temperatures,
and have bright foot points at transition region
temperatures.

Determination of Doppler shifts in Fe {\textsc XII}
line revealed that jets extending from bright points
are blue-shifted by 30 km s$^{-1}$ at maximum.
Taking the projection effect into account,
the observed velocity is consistent with
that of jets expected from chromospheric evaporation
flow caused by magnetic reconnection.
Bright loops at coronal temperature are explained as
loops formed after reconnection between emerging flux
and ambient fields in the coronal hole.
But observed red-shifts near the foot points of loops are
not fully understood in the model.

Collaboration of all three telescopes on board {\it Hinode}
is important for studying the connection between the
photosphere and the corona.
XRT is capable of obtaining a sequence of X-ray images with
high cadence, which is suitable for studying the temporal
variation of the coronal bright points.
The spectro-polarimeter of solar optical Telescope (SOT)
obtains very high resolution transverse and longitudinal
magnetic fields in the photosphere
which are needed for a better understanding of the formation of 
bright points and jets in the corona.

\vspace{12pt}
{\it Hinode} is a Japanese mission developed and launched
by ISAS/JAXA, collaborating with NAOJ as domestic partner,
NASA and STFC (UK) as international partners.
Scientific operation of the {\it Hinode} mission is conducted
by the Hinode science team organized at ISAS/JAXA. This team
mainly consists of scientists from institutes in the partner
countries.
Support for the post-launch operation is provided by JAXA and
NAOJ (Japan), STFC (U.K.), NASA (U.S.A.), ESA, and NSC (Norway).

This work is carried out at the NAOJ {\it Hinode} Science Center,
which is supported by the Grant-in-Aid for Creative Scientific Research
"The Basic Study of Space Weather Prediction"
(17GS0208, Head Investigator: K. Shibata) from the Ministry of
Education, Science, Sports, Technology, and Culture (MEXT) of Japan,
donations from Sun Microsystems, and NAOJ internal funding.

\bibliographystyle{aa}
\bibliography{reference}

\end{document}